\documentclass{article}

\begin{document}

\title{Gamma Ray Bursts statistical properties and limitations on the physical model }

\author{G.S.Bisnovatyi-Kogan}
\maketitle
\begin{abstract}
The present common view about GRB origin is related to cosmology.
There are two evidences in favour of this interpretation. The
first is connected with statistics, the second is based on
measurements of the redshifts in the GRB optical afterglows. Red
shifts in optical afterglows had been observed only in long GRB.
Statistical errors, and possibility of galactic origin of short
GRB is discussed; their connection with Soft Gamma Repeaters (SGR)
is analyzed.
\end{abstract}

\section{Introduction}
Cosmological origin of GRB had been first suggested in \cite{pu75}
soon after their discovery. The present model of cosmological GRB
based on production of gamma quanta from neutrino collisions
$\nu+\bar\nu \rightarrow e^+ + e^-$ was first considered in
\cite{bp87}. The efficiency of transformation of the neutrino flux
$W_{\nu \bar\nu} \sim 6\times 10^{53}$ ergs into gamma quanta was
estimated as $\sim 6\times 10^{-6}$, giving a pulse $W_{\gamma}
\sim 3\times 10^{48}$ ergs. It could explain the cosmological GRB
only at rather narrow pulse beam. In the giant GRB 990123 the
isotropic energy production is very large \cite{kul} in gamma
$W_{\gamma}\approx 2.3 \times 10^{54}$ ergs, and in optics $W_{\bf
opt} \sim 10^{51}$ ergs. Simultaneous strong beaming in gamma and
optical bands is rather unplausible. Strong beaming would modify
the observed smooth optical light curve in presence of a source
rotation. Some problems in GRB interpretation and modelling are
discussed.

\section{Statistics and restrictions to the model}
 BATSE data start to deviate from the uniform
distribution with 3/2 slope at rather large fluences, for which
KONUS data are well defined. Analysis of KONUS data had been done
in \cite{hs90}. Taking into account selection effects, the
resulting value $V/V_{max}=0.45 \pm 0.03$ was obtained. KONUS data
had been obtained in conditions of constant background. Similar
analysis \cite{s99} of BATSE data, obtained in conditions of
substantially variable background, gave resulting $V/V_{max}=0.334
\pm 0.008$. These two results seems to be in contradiction,
because KONUS sensitivity was only 3 times less than that of
BATSE, where deviations from the uniform distribution
$V/V_{max}=0.5$ in BATSE data are still rather large \cite{fm95}.

In presence of a threshold deviations of $V/V_{max}$  from its uniform Euclidean
value 0.5 may be connected with the errors in determination of the
burst peak luminosity or total fluence \cite{bk97}. Such errors
may be connected with spectral differences, variable sensitivity
of detectors for bursts coming from different directions, variable
background. All these reasons lead to underestimation of the
burst luminosity, and decrease the slope of the curve
$\log N (\log S)$. There is no angular correlation between
GRB and sample of any other objects in the universe.

From the energy conservation law it follows $W<M\,c^2$, where $M$ is
a mass of the source.
A proper account of physical laws put much stronger restrictions
to the energy output. Calculations of ns-ns collisions gave energy
output in $(X,\,\gamma)$ region not exceeding $10^{50}$ ergs
\cite{rj98}, and similar results characterize ns-bh collision
\cite{rj99}. Magnetorotational explosion does not give larger energy
output in $(X,\,\gamma)$ region, transforming about 5\% of the
rotational energy into a kinetic one \cite{ar2000}. The problems
with vaguely defined "hypernova" model had been discussed in
\cite{bp98}.

 The largest $\gamma$-ray production efficiency,
close to 100\%$M\,c^2$ may be
expected, if GRB originate from matter-antimatter star
collisions. That arises a problem of antistar creation in the
early universe.

Simultaneous $\gamma,\, X$ and optical observations in GRB,
accompanied by spectral and polarization experiments are very
important. Search of hard $X$-ray lines  and of
annihilation 0.511 keV, line declared by KONUS, remain to be a
puzzle which should be solved. Cosmological GRB explosion in a
dense molecular cloud would lead to a specific optical light curve
\cite{bkt97}, which discovery would reveal conditions in the
region of cosmological GRB explosion. Study of hard $\gamma$
afterglow, similar to the one observed by EGRET \cite{sch} is
expected in a near future.

\section{Short GRB and SGR}
  All afterglows had been measured only for long bursts. It cannot
be excluded that short bursts could have another, may be Galactic
origin. There is also a possibility, that short bursts are
connected with a giant bursts observed in 3 soft gamma repeaters
(from 4 known). At larger distances only giant bursts, appeared as
short GRB could be observed.
   If we accept the present interpretation of SRG, as galactic and
LMC sources at distances 10-50 kpc, than only giant bursts should
be visible in the nearest galaxies as weak (about few $10^{-7}$)
short GRB. Taking into account that Andromeda is $\sim 4$ times more
massive than our Galaxy \cite{vv}, we should expect \cite{bk99}
to see about 10 short GRB
in its direction during the observation time, while
no one was yet observed.
Another large galaxy in the local group of galaxies Maffei
IC 1805 is also more massive than the Galaxy, and short GRB from
it are also expected.

Presently SRG are interpreted as young neutron stars with very
strong magnetic field - "magnetars" \cite{dt}. The estimation of
the distance and, consequently, the luminosity is based on SGR
identification with supernovae remnants (SNR), leading to large
energy losses. This interpretation has several theoretical
objections \cite{bk99}.

1. Hard gamma pulsars observed in 3 SGR have luminosities strongly
exceeding the critical Eddington luminosity. At such luminosity a
strong mass loss should smear out the pulses.

2. Rotational energy losses estimated from the period increase rate
are much smaller than the observed gamma and $X$-ray luminosity
even in a quiescent state. In the magnetar model the energy comes
from the annihilation of magnetic field. Such annihilation should
be accompanied by creation of energetic electrons and
radio-emission. The radio-emission of SGR is very weak, its
discovery is very difficult, and still not firmly established.

3. Giant bursts observed in 3 SGR at present interpretation are
accompanies by a huge energy production, part of which should go
into particle acceleration and kinetic energy outbursts. It should
influence the near-by SNR, and produce a visible changes in radio
and optics, similar to those produced by pulsar glitches in the
Crab nebula, when much smaller amount of energy is released.
No such changes had been reported up to now, probably because they
have not been present there.

Another interpretation of SGR, free of these contradictions needs
a smaller distance to SGR, what is possible if the connection with
SNR would not be confirmed. Note, that all SGR are situated at the
very edge, or even outside of the SNR envelope, requiring very high
1000-3000 km/s speed of the neutron star.
Refusing this connection and suggesting  $\sim 10-30$ times
smaller distance to SGR would remove the upper objections. The
even smaller distances are less probable, because most SGR are
situated in the galactic disk, and so should be situated
 at distances larger than this disk thickness. Existence of one
SGR outside the galactic disk direction could indicate to its
big age during which it could leave the galactic disk.
Discovery of big population of neutron stars in the globular
clusters and in the galactic bulge, as recycled pulsars, indicate to
existence of neutron stars in the whole volume of the Galaxy.

\section{Acknowledgments} This work was partially supported by
RFBR grant 99-02-18180 and INTAS-ESA grant 120. The authors are
grateful to Prof. J.C.Wheeler and the Organizing Committee for
support and hospitality, and to B.Schaefer for discussion.

\end{document}